
\input harvmac.tex
\Title{\vbox{\baselineskip12pt
\hbox to \hsize{\hfill RU-94-32}
\hbox to \hsize{\hfill}
\hbox to \hsize{\hfill}}}
{\vbox{\centerline{Picture-Changing Operators and Space-Time
Supersymmetry}}}
\centerline{Dimitri Polyakov\footnote{$^\dagger$}
{polyakov@pion.rutgers.edu}}
\medskip
\centerline{\it Dept. of Physics and Astronomy}
\centerline{\it Rutgers University,P.O.Box849,Piscataway,NJ 08855}
\vskip .5in
\centerline{\bf Abstract}
\smallskip
We explore geometrical properties of fermionic vertex operators
 for a NSR superstring in order to establish connection between
worldsheet and target space supersymmetries.The mechanism of picture-changing
is obtained as a result of imposing certain constraints
on a world-sheet gauge group of the NSR Superstring Theory.
 We find that  picture-changing operators of different integer
 ghost numbers form a polynomial ring.By using  properties of
the picture-changing formalism, we establish a  relation between
the NSR and GS String theories.We show that, up to picture-changing
transformations,the stress-energy tensor of the $N=1$ NSR superstring
theory can be obtained from the stress-energy tensor of the $N=1$
GS superstring theory in a flat background by a simple field redefinition.
The equations of motion of a GS superstring are shown to be fulfilled in
the NSR operator formalism;they are also shown to be invariant under
$\kappa$-symmetry,in terms of operator products in the NSR theory.
This allows us to derive the space-time
supersymmetry transformation laws for the NSR String Theory.
Then,we explore the properties of the $\kappa$-symmetry in the NSR formalism
and find that it leads to some new relations between bosonic and
fermionic correlation functions.
\Date{December 94}
\vfill\eject
\lref\sherk{F.Gliozzi,J.Sherk,D.Olive,Nucl.Phys.{\bf B122}(1977) 253}
\lref\brink{L.Brink,J.Sherk,J.Schwarz,Nucl.Phys.{\bf B121}(1977) 77}
\lref\fs{D.Friedan,S.Shenker,Phys.Lett. {\bf 160B} (1985) 55}
\lref\fsm{D.Friedan,S.Shenker,E.Martinec,Nucl.Phys.{\bf B271} (1986) 93}
\lref\ver{H.Verlinde,Phys.Lett. {\bf 192B} (1987) 95}
\lref\fried{D.Friedan, Notes on String Theory and 2d CFT,published in
"Workshop on UNIFIED STRING THEORIES",edited by M.Green and D.Gross}
\lref\kni{D.Knizhnik,Usp.Fiz.Nauk 159 (1989) 401}
\lref\ch{G.F.Chapline,N.S.Manton,Phys.Lett {\bf 120B} (1983) 105-109}
\lref\gsw{M.B.Green,J.H.Schwarz,E.Witten,Superstring theory,published by
Cambridge University Press (1987)}
\lref\berk{N.Berkovits,prepr.{\bf KCL-TH-94-5} (April 1994)}
\lref\berkk{N.Berkovits,prepr.{\bf KCL-TH-93-12}(August 1993)}
\lref\vafa{M.Bershadsky et.al. Harvard prepr.{\bf HUPT-A061/92}(Nov.1992)}
\lref\vaffa{N.Berkovits,C.Vafa,prepr.{\bf HUPT-94/A018, KCL-TH-94-12}(July
1994)}
\lref\gates{S.Gates,P.Majumdar et.al.,Phys.Lett.{\bf B214}(1988),26}
\lref\grisaru{M.Grisaru,H.Nishino,D.Zanon,Nucl.Phys.{\bf B314}(1989),363}
\lref\carl{S.Carlip,Nucl.Phys.{\bf B284}(1987),365}
\lref\berkn{N.Berkovits,Nucl.Phys.{\bf B379}(1992),96}
\lref\berkna{N.Berkovits,Nucl.Phys.{\bf B395}(1993),77}
\lref\berknat{N.Berkovits,Nucl.Phys.{\bf B300}(1993),53}
\lref\berknath{N.Berkovits,Nucl.Phys.{\bf B408}(1993),43}
\centerline{\bf 1. Introduction}
In Superstring Theory scattering amplitudes are expressed in terms of
expectation values of vertex operators.For bosons,these operators are
generally of the form $v=f(X,{\partial}X,...{{\partial}^n}X)exp[ikX]$,
where $f$ is some function.The typical example of such an operator is given
by a vector emission
vertex:$V(k,z)=e_{\mu}(k)V^{\mu}(k,z)=exp[ikX]e_{\mu}(k)({\partial}X^{\mu}+i/2(k\psi){{\psi}^{\mu}})$.
In  case of a closed string this expression must also be multiplied by
an antiholomorphic part.
 Construction of the fermionic counterpart of this vector has been
 considered in operator formalism in~refs{\fs, \sherk, \brink} and in the
framework of
 covariant theory~\refs{\fs, \fsm, \ver}These remarkable papers succeded in
constructing scattering amplitudes involving bosons and fermions.
At the same time we do not have clear geometrical reasons explaining
space-time supersymmetry.In this paper we will in particular rederive
the results of the above works in a way which may be more appropriate
for geometrical interpretation.Our basic starting point is the following.
Consider an insertion of
a vertex operator describing the
emission of a spin 1/2 particle at some point z on a worldsheet:
${V_{-1/2}}(z)=
{\bar{{u}^{A}}}(k){{\Sigma}_{A}}(z)exp({-\phi}/2)exp(ikX(z))$,
where $\bar{u}(k)$ is some constant (in a space-time) 10-spinor,$\phi$
is a bosonized superconformal ghost and $\Sigma$ is a spin operator for
matter fields.The index ${A}$ runs from 1 to 32 and the index -1/2
over V  denotes its ghost number.The construction of a fermion emission
vertex is described in~\refs{\fs, \fsm}.
Now,consider the operator product expansion of ${V_{-1/2}}(z)$ with
one of 10 fermionic fields ${\psi_{\mu}}(w)$:
${V_{-1/2}}(z,k,{{\bar{u}}^{A}}(k)){{\psi}_{\mu}}(w)=
{{(z-w)}^{-1/2}}{V_{-1/2}}(w,k,{{\gamma}^{{\mu}AB}}{{\bar{u}}_{B}})$.
It follows from this OPE that the field ${{\psi}_{\mu}}(w)$ changes sign
under the monodromy around z.At the same time the field ${X_{\mu}}(w)$,
its superpartner  does
not change sign under the monodromy.Consequently,the theory does not
have a global supersymmetry.
The same problem appears if we insert a vector vertex (at zero,for
simplicity):
${V_{-1}}(0,k,{e_{\mu}}(k))=exp(-\phi(0)){e_{\mu}(k)}{\psi_{\mu}(0)}exp(ikX(0))$;insertion
of this operator causes the field ${\psi_{\nu}}(z)$ to have a singularity:
${\psi_{\nu}}(z) \sim {{e_{\nu}}(z)\over{z}}$;at the same time
${X_{\nu}}(z) \sim k_{\mu}log(z)$.One way to cure this problem is to
impose certain constraints on gauge parameters.Namely,let us
consider the following locally supersymmetric superstring action:
\eqn\grav{\eqalign{S=1/2\int d^2z
{{(-g){1/2}}}[{g^{ij}}(X){\partial_{i}}{X_{\mu}}{\partial_{j}}{X^{\mu}}+
{\psi_{\mu}}{e^{ij}}(z){\sigma_{j}}{\partial_{i}}{\psi^{\mu}} +\cr
(\chi_{i}{{\sigma}^{j}}{{\sigma}^{i}}{\psi^{\mu}}({{\partial}_{j}}{X_{\mu}}-1/4({\chi_{j}}{\psi_{\mu}}))).}}
here $ {e^{ij}}(z)$ is a zweibein and ${\chi_{i}}(z)$ is a spin 3/2
gravitino field.
The symmetry of this action (apart from general covariance)is given by:

\eqn\grav{\eqalign{{\delta}{X_{\mu}}(z)={\epsilon}(z){{\psi}_{\mu}}(z)
\cr
\delta{{\psi}_{\mu}}(z) =
{{\sigma}^{i}}({{\partial}_{i}}{{X}_{\mu}}(z)-1/2({{\chi}_{i}}(z){{\psi}_{\mu}}(z)))\epsilon(z)
\cr
\delta{{\chi}_{i}}(z)=2{{\nabla}_{i}}{\epsilon}(z) \cr
\delta{{g}_{ij}}=\epsilon(z)({{\sigma}_{i}}{{\chi}_{j}}(z) +
{{\sigma}_{j}}{{\chi}_{i}}(z))}}
Here ${\sigma}_{i}$ are 2d Dirac matrices;$\epsilon(z)$ is a local
infinitezimal fermionic parameter.

To preserve the local supersymmetry in the above example,we must impose
certain constraints on $\epsilon(z)$.Namely,let $z_{1},z_{2}...,z_{N}$
be the points where vertices are
 inserted.The constraints should be
\eqn\lowen{{\epsilon}(z_{1})=0,...,{\epsilon}(z_{N})=0}.
Of course,apart from these requirements,the parameter $\epsilon$ should
have a proper behaviour under the monodromy.
The nature of the requirement (3) is easy to understand.Let us write
down the expression for an N-point scattering amplitude for
fermions~\refs{\kni}:
\eqn\lowen{{A_{N}}(k_{1},...,k_{N})=
{{\int}_{[{z_1},{z_{2}},...{z_N}]}}DXD{\psi}D{\chi}{{{\prod}_{i=1}}^{N}}{\int}d^2 {z_{i}}exp(i{k_{i}}X({z_{i}}))exp(-S)}
Here $[{z_1},{z_2},...{z_N}]$ denotes anti-periodic boundary conditions
that should be imposed on
 the fermions when we move them around the small contours
encircling these points.
The measure and the action in (4) are invariant under local
supersymmetry,however the factors before $exp(-S)$ are not.Therefore,in
order for the amplitude to be locally supersymmetric the supersymmetry
transformation should vanish at the points $z_{1},...,z_{N}$,then
the variation of the integral under the local supersymmetry (2) will be
zero.From this (3) follows.
The imposition of the constraints (3) leads to very important
consequenses.Namely,these conditions effectively reduce the gauge
group (2).As a result  it may no longer be possible
to access the superconformal gauge where the gravitino field may be gauged
away to zero.To investigate this question,it is necessary to consider
the  Faddeev-Popov procedure modified by the constraints (3).
\centerline{\bf 2. Modified FP Procedure}

The usual FP prescription for a gauge group g is defined by
\eqn\lowen{{\Delta}({\chi}){\int}{\delta}({\varphi}({{\chi}^{g}}))Dg=1}
where $\varphi$ is a desired gauge condition for the gravitino field.
Setting $g=I+{\epsilon}$ and imposing the constraints (3) on the gauge group
(2) we can rewrite (5) as
%
\eqn\lowen{{\Delta}({\chi}){\int}{\delta}({\varphi}({{\chi}^{g}})){\delta}({\epsilon}({z_{1}}))...{\delta}({\epsilon}({z_{N}}))D{\epsilon}
=1}
or,equivalently,
%
\eqn\lowen{{\Delta}({\chi}){\int}exp(B{{{\delta}{\varphi}}\over{{\delta}{\epsilon}}}{\epsilon}){\epsilon}({z_{1}})...{\epsilon}({z_{N}})D{\epsilon}DB=1}
where B is some auxiliary fermionic field.
In (7) we also used ${\delta}({\epsilon})={\epsilon}$ since $\epsilon$
is a fermionic variable.
 It is helpful to stress that  it is possible to rewrite any fermionic
functional integral (for instance the one in (7)) as 1 over integral
over some bosonic fields.

For
instance,${\int}exp(B{{{\delta\phi}\over{\delta\epsilon}}}\epsilon)[\epsilon...{\epsilon}]DBD\epsilon={{({\int}D{\beta}D{\gamma}exp(\beta{{\delta\phi}\over{\delta\gamma}}\gamma)[\gamma...\gamma])}^{-1}}$,

where the integration over the fermions $B,\epsilon$ is replaced by that
over the pair of the bosonic fields $\beta,\gamma$.

Thus,instead of expressing ${\Delta}({\chi})$ in terms of the integral (6)
over the pair of fermionic fields $({\epsilon},B)$  we may choose to write
 it as the following integral over the corresponding pair of the bosonic
 fields
${({\beta},{\gamma})}$:
\eqn\lowen{{\Delta}({\chi}) = {\int}D{{\gamma}}D{\beta
}exp({\beta
}{{{\delta}{\varphi}}\over{{\delta}{\gamma}}}{\gamma}){\delta}({{\gamma}}({z_{1}}))...{\delta}({{\gamma}}({z_{N}}))}.
Now let us insert (8) into the partition function:
%
\eqn\grav{\eqalign{Z={\int}D{\chi}D[...]{\Delta}({\chi})exp(-S({\chi},...){\delta}({\varphi}({\chi})))=\cr
{\int}D{{\gamma}}D{\beta}D{\chi}D[...]exp({\beta}{{{\delta}{\varphi}}\over{{\delta}{\gamma}}}{\gamma}){\times}\cr
{\times}{\delta}({\gamma}(z_{1}))...{\delta}({\hat{\gamma}(z_{N})){\delta}({\varphi}({\chi}))exp(-S({\chi},...))}}}
For the sake of brevity,we here suppressed the integration over all
the matter fields except for the gravitino.
The formula (6) shows that,in order to insure the constraints (3) we
have to insert the fields ${\epsilon}(z_{k})$,each field is of a
ghost charge 1.Equivalently,in terms of the bosonic integral (9) we should
insert ${\delta}({\gamma}(z_{k})$.
Analogously,it is the same way to show that
if,instead of imposing constraints (3) we choose to impose
the stronger constraints:
\eqn\grav{\eqalign{{\epsilon}(z_{k})=0 \cr
         {\partial}{\epsilon}(z_{k})=0}}
we have to insert the ghost charge 2 operator
${{\epsilon}}(z_{k}){\partial}{{\epsilon}}(z_{k})$ into the
functional integral.It is straightforwardly generalized to the situation
when,along with ${\epsilon}$ its $(n-1)$ derivatives are zero;we would
have then to insert the ghost charge $n$ operators
${{\epsilon}}...{{\partial}^{n-1}}{{\epsilon}}(z_{k})$.
The corresponding bosonic insertion would be
${\delta}({\gamma}(z_{k})){\delta}({\partial}{\gamma}(z_{k}))...{{\delta}}({{\partial}^{n-1}}{\gamma}(z_{k}))$
It is easy now to see why the superconformal gauge where
${\varphi}({\chi})={\chi}=0$ is
no longer accessible once the constraints (3) are imposed.Indeed,
the functional integrals (6),(9) would  be zero in this case
because of the ghost
charge conservation.Therefore,if we restrict our gauge group (2) with
the constraints (3) we $must$ at the same time relax the gauge condition
on $\chi$.
That is,we should fix the gauge as~\refs{\kni}
\eqn\lowen{{\chi}(z)={{{\sum}_{i=1}}^{M}}{{\chi}_{i}}(z){{\theta}^{i}}},
where ${{\chi}_{i}}$ is a dual basis in the space of
$3/2$-differentials,and ${{\theta}^{i}}$ are the grassmanian parameters
called $supermoduli$.The partition function is shown to be independent
on small variations of the basis ${{\chi}_{i}}$~\refs{\ver}
The number $M$ is equal to the number of
independent $3/2$-differentials;we will compute this number shortly.
A proper choice of the basis in (11) will help us to get the
correct ghost number in the integral (9).Thus,for the gauge constraint
(3) we may choose
\eqn\lowen{{{\chi}_{k}}(z)={\delta}(z-w_{k}),k=1,...M}.
Then,the Faddeev-Popov insertion (5) will write as follows:
%
\eqn\grav{\eqalign{{\Delta}({\chi}){\int}DBD{\epsilon}exp(B{{{\delta}{\varphi}}\over{{\delta}{\chi}}}{\epsilon}){\delta}({\epsilon}(z_{1}))...{\delta}({\epsilon}(z_{N})){\delta}(B(z_{1}))...{\delta}(B(z_{N}))=1
\cr
{\Delta}({\chi})=
{\int}D{\gamma}D{\beta}exp({\beta}{{{\delta}{\varphi}}\over{{\delta}{\epsilon}}}{\gamma}){\delta}({\gamma}(z_{1}))...{\delta}({\gamma}(z_{N})){\delta}({\beta}(w_{1}))...{\delta}({\beta}(w_{M}))}}
Here,as previously,we put ${\varphi}({\chi})={\chi}$.
 Now, let us insert (13) into the partition function.We get:
\eqn\grav{\eqalign{{\int}D{\chi}D[...]D{\beta}D{\gamma}
{{{\prod}_{j=1}}^{M}}{\delta}({\beta}(w_{j}))
exp(-S({\chi},...)){\times}\cr
exp({\int}{\beta}{{{\delta}{\varphi}}\over{{\delta}{\gamma}}}{\gamma}){\delta}({\gamma}(z_{1}))...{\delta}({\gamma}(z_{N})){\delta}({\beta}(w_{1}))...{\delta}({\beta}(w_{M
}))}}
Here the field ${\delta}({\beta})$
is a ghost number $-1$ bosonic field which is to compensate the ghost
number 1 of ${\delta}({\gamma})$.
Note that under the change ${\epsilon}{\rightarrow}{\gamma}$
 the SUSY gauge transformations (3) with the fermionic parameter
$\epsilon$ become BRST gauge transformations with the bosonic ghost
${\gamma}$ as a parameter(
the form of these transformations remains the same as in (3),with
$\epsilon$ replaced by $\gamma$).
Now it follows from (3) that,since
${{\delta}_{SUSY}}{{\chi}_{i}}(z)=2{{\nabla}_{i}}{\epsilon}(z)$, the
corresponding BRST gauge transformation is
\eqn\lowen{{{\delta}_{BRST}}{{\chi}_{i}}(z)=2{{\nabla}_{i}}{\gamma}(z)}
and therefore
%
\eqn\lowen{{{{\delta}{\varphi}({{\chi}_{i}})\over{{\delta}{\epsilon}}}}={{\delta}{\varphi}({{\chi}_{i}})\over{{\delta}{\gamma}}}=2{{\nabla}_{i}}}
We now have to integrate over the gauge field ${\chi}$.
One problem still remains with the bosonic ghost action in
(14).Namely,this action is not BRST invariant.One may check that
 ${Q_{BRST}}{S_{gh}}({\beta},{\gamma},b,c){\neq}0$,where
\eqn\grav{\eqalign{{S_{gh} =
{\int}{d^2}z[-{\beta}{\nabla}{\gamma}+b{\nabla}c]}\cr
{Q_{BRST}}={\oint}{d^2}z[{\gamma}({G_{m}}+1/2{G_{ghost}})+c({T_{matter}}+1/2{T_{ghost}})]}}
Here $G_{matter}$,$G_{gh}$ are  the supercurrents for matter and ghost
fields respectively;and ${{T}_{m}}$,${{T_{gh}}}$ are stress-energy
tensors.In order to restore the BRST invariance we must add the term
${\sim}{\chi}{G_{gh}}$ to the ghost action;this term is analogous to
${\sim}{\chi}{G_{gh}}$ we have already got in the matter part in order
to insure the local supersymmetry (3).
Now we are prepared to perform the integration over the gravitino field.
Due to the gauge fixing condition (11) it will simply be an integral
over M supermoduli ${{\theta}^{i}}$;the whole dependence
 of the modified action on the gravirino field
in (14) contains in the piece
${\sim}{\int}{d^2}z{\chi}({{G}_{m}}+{{G}_{gh}})$;
therefore,by using (11),expanding
the exponent and performing the integral over ${\theta}$'s we obtain the
following insertion into the partition function (14):
\eqn\lowen{{{\prod}_{k=1}^{M}}[{{G}_{m}}(w_{k})+{{G}_{gh}}(w_{k})]}

Putting this together with the insertion in (14)that comes from FP determinant
with relaxed gauge conditions we find that the total insertion depending
on the points $w_{k}$ chosen for the basic vectors in (11) will be:
\eqn\lowen{{{{\prod_{k=1}^M}}}{{\Gamma}_{1}}(w_{k})}
where
%
\eqn\lowen{{{\Gamma}_{1}}(w_{k})={\delta}({\beta}(w_{k}))({{G}_{m}}(w_{k})+{{G}_{gh}}(w_{k}))}
The operator ${{\Gamma}_{1}}$ is known as a picture-changing
operator~\refs{\ver}
 Bosonization formulas for ${\beta}$,${\gamma}$ ghost system are the
 following:
\eqn\grav{\eqalign{{\gamma}(z)=exp({\phi}(z)-{\kappa}(z)) \cr
{\beta}(z)=exp({\kappa}(z)-{\phi}(z)){\partial}{\kappa}(z)\cr
{\delta}({\beta}(z))=exp({\phi}(z))\cr
<{\kappa}(z){\kappa}(w)> = - <{\phi}(z){\phi}(w)> = log(z-w)}}
Bosonization rules for ${\delta}(\beta)$ are explained in~\refs{\fried}.
There exists a simple way to rederive them through relations
between bosonic and fermionic correlators.
The fermionic fields $\epsilon$ and $B$ may be bosonized as
$\epsilon=exp[i{\hat{\phi}}]$;$B=exp[-i{\hat{\phi}}]$ with
$<\hat{\phi}(z)\hat{\phi}(w)>=ln(z-w)$.On the other
hand,by using $\delta(\epsilon)=\epsilon$,$\delta(B)=B$,we can write:

$<\epsilon(z_{1})...\epsilon(z_{N})B(w_{1})...B(w_{N})>=<\delta(\epsilon(z_{1}))...\delta(\epsilon(z_{N}))\delta(B(w_{1}))...\delta(B(w_{N})>$=

$={{({<\delta(\gamma(z_{1}))...\delta(\beta(z_{1}))...\delta(\beta(z_{N}))>})}^{-1}}$.

Substituting the bosonization formulas for $\epsilon$ and $B$ and recalling
that

$<{\prod_{i}}exp[{\alpha_{i}}\phi(z_{i})]>=exp{\sum_{i,j}}{\alpha_{i}}{\alpha_{j}}<\phi(z_{i})\phi(z_{j})>$
we find that the bosonization formulas (21) are correct with $\phi
=i{\hat{\phi}}$.
{}From here it is clear why the correlator of 2
$\phi$'s should have the "wrong" sign.
Note that the conformal dimension of the field $exp({\alpha}{\phi})$
is~\refs{\gsw}
$-{{{\alpha}^2}\over{2}}-{\alpha}$.
Therefore the expression for the picture-changing operator (20) in terms
of free fields is:
%
\eqn\lowen{{{\Gamma}_{1}}(z)=exp({\phi}(z))[-1/2{{\psi}^{\mu}}{\partial}{X_{\mu}}(z)-1/2b{\gamma}(z)+c{\partial}{\beta}(z)+3/2{\beta}{\partial}c(z)]}
This operator has zero conformal dimension.
Of course, a proper normal ordering must be done in the formula (22).

\centerline{\bf 3.Counting the number of parameters}
Now we have to find the number M of the holomorphic differentials
${{\eta}^{\beta}}(z)$ which is equal to the number of their dual
${{\chi}_{\alpha}}$(z) in (11).The duality means that
%
\eqn\lowen{{\int}{d^2}z{{\eta}_{\beta}}{{\chi}_{\alpha}}={{\delta}_{\alpha\beta}}}
In case of an amplitude on a sphere this number has been already
computed in~\refs{\kni} and has been shown to be equal to ${{N-4}\over{2}}$ for
an N-point fermionic
amplitude.The gauge constraints (3) were discussed there.
However,we would like to count the number of the basic vectors in a case
when other  constraints (like (10) or stronger) are imposed.
Our approach  mostly  repeats~\refs{\kni}.
Let us consider the most general case of constraining the gauge (2).
{}From now on,unless stated otherwise we will restrict ourselves to
the problem on a sphere.
Let's consider the points ${z_{i}}$,$i=1,...N$ where the vertices are inserted.
Suppose that,at the point ${z_{i}}$ we have ${\epsilon}(z_{i})$ together
with its first ${k}_{i}$ derivatives vanishing.
Then, the basis in the space of holomorphic 3/2 differentials may be chosen as
%
\eqn\lowen{{{\eta}^{\beta}}={{z}^{\beta}}{\prod_{i=1}^N}{{(z-z_{i})}^{-1/2-{k_{i}}}}}
The number of independent differentials is therefore equal to the number
of all admissible $\beta$'s.The condition of holomorphy for 3/2
differentials requires that (24) goes to zero at infinity no slower
than ${1\over{z^3}}$.
This leads to the condition
\eqn\lowen{{\beta}-N/2-{\sum_{i=1}^N}{k_{i}}{\leq} -3}
Hence it follows that there are $M={{N-4}\over{2}}+{\sum_{i=1}^N}{k_{i}}$
independent holomorphic 3/2 differentials corresponding to (24).

\centerline{\bf 4.Further gauge constraints}
All the above arguments may be repeated if we choose to impose the
constraints (10) instead of (3) on the gauge group(2).
In this case,we will have to make the insertion of a ghost charge 2
FP determinant,as explained previously.In order to compensate this
ghost charge we again will have to relax the gauge condition for the
gravitino field.The proper choice of  the dual basis ${{\chi}_{i}}$
 will now be
\eqn\lowen{{{\chi}_{i}}(z)={\partial}[{\delta}(z-{{w}_{i}})]}
and the gauge fixed gravitino field will be
%
\eqn\lowen{{\sum_{i=1}^{{3n-4}\over{2}}}{\partial}[{\delta}(z-{{w}_{i}})]{{\theta}^{i}}}.
The assosiated insertion will be
\eqn\lowen{{\prod_{j=1}^{{3N-4}\over{2}}}{{\Gamma}_{2}}({w}_{j})}
where
\eqn\lowen{{{\Gamma}_{2}}(w_{k})=
:{\delta}({\beta}){\delta}({\partial}{\beta})G{\partial}G:(w_{k})}

here G is a full supercurrent:$G=G_{m}+G_{gh}$.
Bosonizing according to (21) we get:
\eqn\lowen{{{\Gamma}_{2}}(w_{k})= :exp(2{\phi})G{\partial}G:(w_{k})}
This is $another$ picture-changing operator which has a ghost
number 2.Again,it has a zero conformal dimension.
One may easily check  the following obvious generalization for the last
formula.  That is, requiring ${\epsilon}$ with its $n-1$ derivatives to vanish
 in (2) leads to insertion of the following
 picture-changing  operator of a ghost charge $n$:
\eqn\lowen{{{\Gamma}_{n}}= :exp(n{\phi})G{\partial}G...{{\partial}^{n-1}}G:}
All the picture-changing operators (31) have zero conformal dimensions.

\centerline{\bf 5. Properties of the Picture-Changing Operators}

First of all,let us write down the normal ordered expression for the
picture-changing operator (20).We have:
%
\eqn\lowen{:{{\Gamma}_{1}}:=exp({\phi})(G_{m}+G_{gh})=-1/2exp({\phi}){{\psi}_{\mu}}{\partial}{X^{\mu}}-1/2exp(2{\phi}-{\kappa}){\partial}{\phi}b+exp({\kappa}){\partial}{\kappa}c}
Let us now calculate the zero order term in
 the operator product of ${{\Gamma}_{1}}$ with the
fermionic vertex operator $V_{-1/2}$
We get:
\eqn\grav{\eqalign{:exp({\phi}(z)){{\psi}_{\mu}}{\partial}{X^{\mu}}(z)
::{{\bar{u}}^{A}}(k){{\Sigma}_{A}}exp({-\phi}(w))exp(ikX(w)):{\sim}\cr
{\sim}...+{(z-w)^{0}}{{{\gamma}^{\mu}}_{AB}}{{\bar{u}}^A}(k){{\Sigma}_{B}}[{\partial}{X^{\mu}}+i/4(k{\psi}){{\psi}^{\mu}}]exp(ikX(w));\cr
:exp({\kappa}){\partial}{\kappa}c{{\bar{u}}^A}{{\Sigma}_{A}}exp(-1/2{\phi})exp(ikX):{\sim}\cr
{\sim}exp({\kappa}-1/2{\phi}){{\bar{u}}^A}{{\Sigma}_{A}}c
 exp(ikX) ;\cr
:exp(2\phi-\kappa)\partial{\phi}b{{\bar
u}^A}{{\Sigma}_{A}}exp(-1/2\phi)exp[ikX]:=exp(3/2\phi-\kappa)b{{\bar
u}^A}{{\Sigma}_{A}}exp[ikX];\cr
:{{\Gamma}_{1}}{V_{-1/2}}({{\bar{u}}^A}(k),z):{\sim}{V_{1/2}}({{{\gamma}^{\mu}}_{AB}}{u^{B}},w)+cexp({\kappa}){\partial}{\kappa}{V_{-1/2}}(w)+\cr
+1/2bexp(3/2\phi-\kappa){{\bar u}^A}{{\Sigma}_{A}}exp[ikX]}}
Here
%
\eqn\lowen{{V_{1/2}}={{{\gamma}^{\mu}}_{AB}}{\bar{u}^A}{{\Sigma}^{B}}[{\partial}{X_{\mu}}+i/4(k{\psi}){{\psi}_{\mu}}]exp(ikX)}
The terms proportional to the ghost fields $b$and $c$ in the product
$:{{\Gamma}_{1}}{V_{-1/2}}:$ may be omitted since it does not contribute
to any correlation function~\refs{\fsm}
{}.
The operator $V_{1/2}$ is a well known fermionic vertex operator in the
 picture of ghost number 1/2
Now let us consider a construction of the
fermionic vertex in the picture of a ghost number -3/2.The obvious candidate
(the space-time spinor whose conformal dimension is 1) is:
\eqn\lowen{{V_{-3/2}}={{\bar{u}}^A}{{\Sigma}_{A}}exp(-3/2{\phi})exp[ikX]}
The normal ordered product of this operator with ${\Gamma}_{1}$ will
give us

$:{{\Gamma}_{1}}{V_{-3/2}}:{\sim}{({\gamma}k)_AB}{{\bar{u(k)}}^A}{V_{-1/2}^B}=0,$
because of the on-shell condition.

The reason for that is the BRST non-invariance of $V_{-3/2}(k)$.
In order to make it BRST invariant one should add to it the term equal
to

${{\tilde{V}}_{-3/2}(k)}=1/2{{{\bar{u}}(k)}^A}{\Sigma_A}exp[\kappa-5/2\phi]{\partial^2}c{\times}exp[ikX]$.
As one may easily check,${{\tilde{V}}_{-3/2}}(k)$ has the conformal
dimension 1,and its BRST variation compensates that of $V_{-3/2}$;
also $:{\Gamma_1}{{\tilde{V}}_{-3/2}}(k):{\sim}V_{-1/2}(k)$,up to
terms that do not contribute to correlation functions.

Similarly,by acting on $V_{1/2}$ with ${\Gamma}_{1}$ we obtain
fermionic vertex in the $3/2$ picture.However, this procedure
contains the subtlety that  will be discussed below.
The same is true for boson emission vertices.
For example, let us consider the emission of a vector boson:
${V_{-1}}(k)={e_{\mu}}(k)exp(-{\phi}){{\psi}^{\mu}}exp[ikX]$.
 The product with ${\Gamma}_{1}$ gives
\eqn\lowen{:{{\Gamma}_{1}}{V_{-1}(k)}:{\sim}{e_{\mu}}(k)[{\partial}{X^{\mu}}+i/2(k{\psi}){{\psi}^{\mu}}]exp[ikX]+{\beta}c{e_{\mu}}{{\psi}^{\mu}}exp[ikX]}
Again,the term $\sim$ c is irrelevant since not  contributing to any
correlation function.
Therefore
\eqn\lowen{:{{\Gamma}_{1}}V_{-1}:{\sim}V_{0}}
where
${V_{0}}={e_{\mu}}[{\partial}{X^{\mu}}+i/2(k{\psi}){{\psi}^{\mu}}]exp[ikX]$

In general,if we have a vertex operator of a ghost number m (m is
integer for bosons and half-integer for fermions),its product with the
picture-changing operator (20) will give us the operator describing the
emission of the same particle with momentum $k$ but of a ghost number $m-1$.
By applying BRST charge (17) to both left and right hand side of (36) it
is easy to see that the operator ${{\Gamma}_{1}}$ is BRST invariant
since both rhs of (36) and $V_{-1}$ are invariant under BRST.
There is another very important property possessed by the operator
${{\Gamma}_{1}}$.That is, if we insert ${{\Gamma}_{1}}$(a) into any correlation
function:
$<{{\Gamma}_{1}}(w){V_{1}}(z_{1})...{V_{N}}(z_{N})>$,this correlation
function will be independent on the point of insertion $w$.
This means that the operator ${{\Gamma}_{1}}$ can be moved freely from
$w$ to any other point in a correlator
 - the correlation function will remain unchanged.This leads
to important identities  for  correlation functions
For example,let us consider the three-point correlation function on a
sphere:
$<V_{0}(z_{1})V_{-1}(z_{2})V_{-1}(z_{3})>$.
We have:
%
\eqn\grav{\eqalign{<V_{0}(z_{1})V_{-1}(z_{2})V_{-1}(z_{3})>=<{{\Gamma}_{1}}({z_{1}}-{\epsilon}){V_{-1}}(z_{1})V_{-1}(z_{2})V_{-1}(z_{3})>=\cr
=<V_{-1}(z_{1}){{\Gamma}_{1}}(z_{1}+{\epsilon})V_{-1}(z_{2})V_{-1}(z_{3})>=
<V_{-1}(z_{1})V_{-1}(z_{2}){{\Gamma}_{1}}(z_{2}+{\epsilon})V_{-1}(z_{3})>\cr=
=<V_{-1}(z_{1})V_{0}(z_{2})V_{-1}(z_{3})>=
<V_{0}V_{0}V_{-2}>=<V_{1}V_{-1}V_{-2}>=...etc}}.
In the similar way we can  show that the operator
${{\Gamma}_{2}}$ increasing the ghost number of vertices by 2
has the properties that are the same as that of ${{\Gamma}_{1}}$.
The same is true for any picture-changing operator ${{\Gamma}_{n}}$.
However, the separate manipulations with  ${{\Gamma}_{1}}$,
${{\Gamma}_{2}}$,... do not give us the
the full set of existing identities between the correlation functions.
We need to establish connection between different $\Gamma$'s.
The problem here is the following.There are,for instance, two ways
 to change the ghost number of $V_{m}$ by 2.
We can make it by  acting on it either twice with ${{\Gamma}_{1}}$ or
once with ${{\Gamma}_{2}}$.
The natural question is whether the results will be the same
(up to maybe BRST trivial terms).
For example, acting on $V_{-3/2}$ twice with ${{\Gamma}_{1}}$ is,up to
BRST trivial term,the same as acting on it once with  ${{\Gamma}_{2}}$
-in both cases we obtain $V_{1/2}$,up to terms that do not contribute to
a correlation function.
The question is that whether this is true for an arbitrary vertex operator
$V_{m}$ and the picture-changing operators ${{\Gamma}_{k}}$ , ${{\Gamma}_{l}}$
and ${{\Gamma}_{k+l}}$.
Let us check this for two ${\Gamma_{1}}$'s.
We need to compare the expressions for $:{{\Gamma}_{1}}{{\Gamma}_{1}}:$
and ${{\Gamma}_{2}}$ obtained as a result of a normal ordering.

The normal ordering of ${{\Gamma}_{1}}{{\Gamma}_{1}}$ gives:
\eqn\grav{\eqalign{:{{\Gamma}_{1}}{{\Gamma}_{1}}:=1/4exp(2{\phi})[{G_{m}}{\partial}{{G}_{m}}+
{c_{m}}{{{P}^{4}}_{{\phi}}}+({\partial}{\phi}{\partial}{\phi}-{{\partial}^2}{\phi})T_{m}]-\cr
-1/2exp(3{\phi}-{\kappa})[{\partial}G_{m}{\partial}{\phi}b
-
G_{m}[{\partial}({\partial}{\phi}b)+{\partial}{\phi}b({\partial}{\kappa}-{\partial}{\phi})]]+\cr
+1/4exp(4{\phi}-2{\kappa}-2{\sigma}){\sum_{i,j<4}}{{{P}^i}_{\phi}}{{{P}^j}_{\sigma}}{{{P}^{4-i-j}}_{2{\phi}-{\kappa}}}+\cr
+1/4exp(2\phi)[c_{gh}{{{P}^4}_{\phi}}+{T_{gh}}{{{P}^2}_{\phi}}]+1/2exp(2{\kappa}+2{\sigma})}}
Here $c_{m}$ and $c_{gh}$ are the matter and ghost central
charges,$\sigma$ is the bosonized fermionic ghost :
$c=exp(\sigma)$,$b=exp(-\sigma)$;
${P^i}_{\phi}$ is the polynomial in the derivatives of ${\phi}$ which
has conformal dimension $i$ and is determined by the OPE of
$exp(\phi)$ with itself:
%
\eqn\lowen{exp(\phi(z))exp(\phi(w)){\sim}exp(2\phi(w))[{1\over{z-w}}+{\sum_i}{{P^i}_{\phi}}{(z-w)^{i-1}}]}
For example,${{P^2}_{\phi}}=\partial\phi\partial\phi-{{\partial}^2}{\phi}$.
The polynomials
${{P^j}_{\sigma}}$and${{P^{4-i-j}}_{2\phi-\kappa}}$ are defined in the
same way.
The normal ordering of ${{\Gamma}_{2}}$ gives:
%
\eqn\lowen{:{{\Gamma}_{2}}:=:{{\Gamma}_{1}}{{\Gamma}_{1}}:-(\partial\phi\partial\phi-{{\partial}^2}\phi)(T_{m}+T_{gh})-1/4exp(2\phi)[(c_{m}+c_{gh}){{{P}^4}_{\phi}}]}
Since $c_{m}+c_{gh}=0$ we can write the last equation as
\eqn\lowen{:{{\Gamma}_{2}}:=:{{\Gamma}_{1}}{{\Gamma}_{1}}:-{\lbrace}{Q_{BRST}},(\partial\phi\partial\phi-{{\partial}^2}{\phi})b{\rbrace}}
Here we used
%
\eqn\grav{\eqalign{{\lbrace}{Q_{BRST}},(\partial\phi\partial\phi-{{\partial}^2}{\phi})exp(2\phi){\rbrace}=0\cr
{\lbrace}{{Q}_{BRST}},b{\rbrace}=T_{m}+T_{gh}}}
Therefore,$:{{\Gamma}_2}:=:{{\Gamma}_{1}{\Gamma}_{1}}:$ up to BRST
trivial term.
Analogously,one may show that

\eqn\grav{\eqalign{:{{\Gamma}_{m}}{{\Gamma}_{n}}:=:{{\Gamma}_{m+n}}:+
{\lbrace}{Q_{BRST}},...{\rbrace};\cr
:{{\Gamma}_{n_{1}}}...{{\Gamma}_{n_{k}}}:=:{{\Gamma}_{{n_{1}}+...+{n_{k}}}}+{\lbrace}{Q_{BRST}},...{\rbrace}:.}}
The straightforward proof of (44) is rather cumbersome;let us first
prove it for ${{\Gamma}_{m}}$ and ${{\Gamma}_{1}}$.
Rather lengthy calculation shows that in this case
%
\eqn\lowen{:{{\Gamma}_{m}}{{\Gamma}_1}:=:{{\Gamma}_{m+1}}:+{\lbrace}{Q_{BRST}},{\sum_{l=1}^m}{\sum_{k=1}^l}{{{(-1)^k}(l+k-1)!}\over{l!}}{P_{(m{\phi},{\phi})}^{m+k}}{{\partial}^{l-k}}b{\rbrace}}
Here ${P_{(m{\phi},{\phi})}^{m+k}}$ is again the polynomial in
the derivatives of ${\phi}$ which has the conformal dimension $m+k$
and  is obtained in the process of the OPE of $exp(m{\phi})$
 with $exp({\phi})$.
Then,(44) may be proved by induction.
The formulas (42)-(44)
mean that the picture-changing operators (31) form
an infinite-dimensional polynomial ring in the sense described above.
In the language of gauge constraints this means that there are many
equivalent ways to impose constraints on (2).
For example,(42) implies that the constraint that requires
$\epsilon$ to vanish at
two different points is equivalent to requiring both $\epsilon$ and
$\partial\epsilon$  to be zero at the same point on a world sheet.

It follows from (38)-(44) that any correlation function
$<V^1_{i1}(k_{1})...V^N_{iN}(k_{N})>$,($i_{k}$ denotes a ghost number of
a vertex $V^k(p_{k}))$is independent upon a distribution of ghost numbers
among the vertices, and the only required condition is
\eqn\lowen{{\sum_{k=1}^N}i_{k}=2(g-1)}
where g is a genus of the surface on which the correlation function is
computed,i.e. the total ghost number must compensate the ghost
number anomaly on the surface.

For a sphere
\eqn\lowen{{\sum_{k=1}^N}i_{k}= -2}
In other words,due to (44)it does not matter which pictures
 we choose for  vertex operators in the correlator;
we only need to fix the total ghost number.
Also because of (44) the fermionic vertex operator with an
arbitrary ghost numper $n-1/2$ is now defined unambiguously:
\eqn\lowen{V_{n-1/2}=:{{\Gamma}_{n}}V_{-1/2}:}
The same is true for the vertex describing the emission of massless
vector bosons.
\centerline{\bf 6.Fermionic Background}
The relation (44) between the picture-changing operators allows us to
construct the consistent perturbation theory for a NSR superstring
in order to derive the superstring corrections to the classical
equations of motion for the supergravity.The superstring corrections
were discussed in~\refs{\gates, \grisaru} by using semi-light-cone
gauge fixing~\refs{\carl} though the results were not yet complete.
In order to obtain the low-energy effective
 action  in the field theory limit of the superstring theory we need
to perturb the superstring action in a flat background
\eqn\lowen{S_{0}=\int
{d^2}{z}({\partial}{X^{\mu}}{\bar{\partial}}{X_{\mu}}+{{\psi}^{\mu}}{\bar{\partial}}{{\psi}_{\mu}})}
with the sum of massless background fields multiplied by corresponding vertex
operators.For example, the term describing

the perturbation by a graviton is
${\sim}{\int}d^2{z}G_{\mu\nu}(X){\partial}{{X}^{\mu}}{\partial}{{X}^{\nu}}$,

where $G_{\mu\nu}(X)={{\eta}_{\mu\nu}}+t_{\mu\nu}(X)$,
${{\eta}_{\mu\nu}}$ is a flat space-time metric and $t_{\mu\nu}$ is its
small perturbation.
Expanding the partition function of
the perturbed action into series in ${t_{\mu\nu}}$,extracting the
cutoff dependence, we obtain the graviton's $\beta$-function (which
is the Ricci tensor in the zero loop approximation).Then,the equations
of motion $\beta (G_{\mu\nu})=0$ define the low-energy gravitational action .
In order to obtain the low-energy action for the supergravity (which has
been calculated in the first orders in~\refs{\ch}
we also need to perturb  $S_{0}$ with a 10-dimensional gravitino field
times the fermionic vertex operator:
\eqn\lowen{S_{1}=\int{d^2}{z}[G_{\mu\nu}(X)\partial{X^{\mu}}{\bar{\partial}}{X^{\nu}}+{\chi}_{\mu}^{(10)}(X){V_{\alpha}^{\mu}}]}
where $V_{\alpha}^{\mu}$ is a fermion vertex operator $V_{\alpha}$
multiplied by ${\bar{\partial}}X^{\mu}$;
the index $\alpha$ depends on which picture we choose for the
fermionic vertex.$V_{\alpha}^{\mu}$ is the vertex operator describing
the emission of a spin 3/2 fermion.
However, there are several difficulties with the formula (50).
First of all, it is not
clear which picture  should be chosen for the fermionic vertex operator
$V_{\alpha}$ in (50).If,for instance, we choose $\alpha=-1/2$,so that
\eqn\lowen{{V_{-1/2}^{\mu}}={{\bar
u}^{A}}{\Sigma_{A}}exp(-1/2\phi){\bar{\partial}}{X^{\mu}}{exp[ikX]}}
it is not clear how to develop a perturbation theory for the background (50)
Indeed,the only non-vanishing correlation function of ${V_{-1/2}^\mu}$'s
is four-point,due to (47).The same problem appears if, instead of
$\alpha=-1/2$ picture we choose any other half-integer value of $\alpha$.
We need to modify the perturbed partition function in order to insure
the correct total ghost number(47) in every term of the expansion.
In order to obtain this,the nth term in the expansion of (50) which is
proportional to ${{\chi_{\mu}^{10}}^n}$ must be multiplied by
${{\Gamma}_{{{n-4}\over{2}}}}$.This goal will be achieved if we write down
the perturbed partition function as follows:
%
\eqn\lowen{{Z_{pert}}={\int}DXD{\psi}{1\over{1-{{\Gamma}_{1}}}}exp[{G_{\mu\nu}}(X){\partial}{X^{\mu}}{\bar{\partial}}{X^{\nu}}+{{\chi}_{\mu}^{10}}V_{-1/2}^{\mu}]}
Due to (44) and (47) the factor $1\over{1-{{\Gamma}_{1}}}$ inserted
 in (52) insures that all the terms in
expansion of the exponent in (52) automatically have the correct ghost number.
We conclude that,in order to obtain correct perturbed partition function
for NSR superstring one must insert
the function $f({{\Gamma}_{1}})={1\over{1-{{\Gamma}_{1}}}}$ of the picture
changing operator  in the measure of integration.
\centerline{\bf 7.Relation to Green-Schwartz superstring.
}
The algebra (44) describes the operators that change ghost numbers
of bosonic or fermionic vertices by integer values;the question arises
whether it is possible to extend $m$ and $n$ in (44) to half-integer
values.Operators changing ghost numbers by half-integer values
transform bosonic vertices into fermionic  and vice versa.
Polarisation spinors of the fermions obtained this way
will be the expressed in terms of polarisation vectors of bosons
multiplied by space-time gamma-matrices or their combination.
Therefore,
this extention may be justified only when the theory possesses the
space-time supersymmetry,i.e. there exists a symmetry between
bosonic and fermionic vertex operators.The problem is therefore to
find the explicit form of the local "picture-changing" operator
${\Gamma}_{1/2}$(other half-integer "picture-changing" operators can
be obtained by the multiplication with integer  $\Gamma$'s.Given that
the expression for ${\Gamma}_{1/2}$ is found,the question that arises is
whether it is possible to "move" it  inside the correlator like it
is done in (38).If the answer is positive,we can establish the relations
between bosonic and fermionic amplitudes that do not follow directly
from space-time supersymmetry.
 The operator ${\Gamma}_{1/2}$ should be proportiponal to $exp[1/2\phi]$
and be of zero conformal dimension.The only local operator that
satisfies
 these
properties is
\eqn\lowen{{{\Gamma_{1/2}}}=exp[1/2\phi]\Sigma}

However,this operator cannot be the one we are looking for since
${{\Gamma}_{1/2}}{{\Gamma}_{1/2}}{\neq}{{\Gamma}_{1}}$ and therefore
cannot be used for the extention of the algebra (44);also by direct
computation one can show that, for a vertex operator $V_{n}$
the identity ${{\Gamma}_{1/2}}V_{n}(k)=V_{n+1/2}(k)$
does not hold in general case.Also, this operator is not BRST invariant.
However,some of its  properties that will be helpful in order
to discuss  the questions formulated above.
We will find out shortly
that it has a  relevance to the Green-Schwartz
(GS) superstring theory.Let us now recall some basic facts from this theory.
The action of a Green-Schwarz superstring  is given by~\refs{\gsw}:
\eqn\grav{\eqalign{S=S_{1}+S_{2},\cr
S_{1}=-{1\over{2\pi}}{\int}{d^2}zg^{1/2}g^{\alpha\beta}{{\Pi}_{\alpha}}{{\Pi}_{\beta}},\cr
S_{2}={1\over{\pi}}\int{d^2}z[-i{{\epsilon}^{\alpha\beta}}{\partial_{\alpha}}X^{\mu}\theta{{\gamma}_{\mu}}{\partial_{\beta}}\theta]}}
Here
%
\eqn\lowen{{{\Pi}^{\mu}_{\alpha}}={\partial_{\alpha}}X^{\mu}-i\theta{\gamma^{\mu}}{\partial_{\alpha}}\theta}
Here $\theta$ is a 10-dimensional spinor and a world-sheet scalar.
For simplicity,we here consider a heterotic string.
The global space-time supersymmetry  for this theory is written as
follows~\refs{\gsw}:
\eqn\grav{\eqalign{\delta{\theta^A}={\epsilon^A}\cr
{\delta}X^{\mu}=i{\epsilon^A}{{\gamma^{\mu}}_{AB}}{\theta^{B}}\cr
{\delta}e=0}}
Here $A=1,...2^4$is a spinor index.
Apart from this global space-time supersymmetry the action (54) also
possesses the following local fermionic $\kappa$-symmetry given by~\refs{\gsw}:
\eqn\grav{\eqalign{{\delta}{{\theta}^A}=2i{{\gamma}^{AB}}{\Pi}{{\kappa}_{B}}\cr
{\delta}{X^{\mu}}=i{\theta^A}{{\Gamma}_{AB}}{\delta}{{\theta}^B}\cr
{\delta}({g^{1/2}}g^{\alpha\beta})=-16{g^{1/2}}({P^{\alpha\gamma}}{\kappa^{\beta}}{{\partial}_{\gamma}}{\theta})}}
where
$P^{\alpha\gamma}=1/2(g^{\alpha\gamma}-{g^{-1/2}}{\epsilon^{\alpha\gamma}})$.
$\kappa$-symmetry insures that half of the components of
$\theta$ are decoupled from the theory - and therefore the number of
these components is proper.
The equations of motion in the conformal gauge are:
\eqn\grav{\eqalign{{\gamma_{\mu}}{\Pi_z^{\mu}}{\bar{\partial}}\theta=0\cr
{\bar \partial}[\partial{X^\mu}+i\theta{\gamma^\mu}\partial\theta]=0;}}
 2 other equations are obtained by complex conjugation.
The momenta conjugate to the X and $\theta$ coordinates are given by
${\pi_{X}}=\Pi$,${\pi_{\theta}}=i\gamma\Pi\theta$.
Because of this phase-space constraint the quantization procedure
becomes very cumbersome in covariant gauges and the problem of covariant
quantization of the Green-Schwarz superstring has not yet been
successfully solved.
It was explained in~\refs{\berkn, \berkna, \berknat, \berknath} how
to perform free field quantization
for the $N=2$ GS superstring
by using a BRST charge constructed out of a stress-energy tensor with $c=6$.
However,in the formalism developed in these papers the amplitudes were
not manifestly Lorentz-covariant in 10 dimensions;rather,they only
were evidently covariant under $SO(3,1)$ subgroup of the super-Poincare
transformations.
Therefore,the problem of the covariant quantization of the 10
dimensional GS superstring in a flat background remains unsolved.

A possible approach to this problem is to establish direct relation between
GS and NSR variables;for the latter theory the covariant quantization
procedures are well-known.The idea is~\refs{\berk} to identify $\theta$ and
${{\Gamma}_{1/2}}=exp[1/2\phi]\Sigma]$ from the NSR theory by using the
fact that space-time and world-sheet transformation properties of these
operators are identical:they both are world-sheet scalars and space-time
spinors.Therefore,the identification we have to prove to be correct is
\eqn\lowen{\theta=exp[1/2\phi]\Sigma}
We need to prove that,under this identification,the global supersymmetry
transformations (56) and the equations of motion (58) are still fulfilled.
Let us check the supersymmetry first.In the NSR theory,the space-time
supersymmetry generator is given by
\eqn\lowen{Q_={\epsilon^A}{\oint}{{dz}\over{2i\pi}}exp[-1/2\phi]{\Sigma_{A}}.}
Therefore,by using (59),(60) and performing a product
expansion,  we have
%
\eqn\grav{\eqalign{{{\delta}_{SUSY}}{\theta_{A}}={\epsilon^B}{\oint}{{dz}\over{2i\pi}}exp[-1/2\phi(z)]{\Sigma_{B}}(z)exp[1/2\phi(w)]{\Sigma_{B}}(w)=\cr
={\epsilon^B}{\oint}{{dz}\over{2i\pi}}[{1\over{z-w}}{\epsilon_{BA}}+...]={\epsilon^B}{\epsilon_{AB}}}}.
We can redefine the definition (59) of Q by multiplying it by the constant
antisymmetric tensor${\epsilon^{AB}}$,then we will return to the
transformation law (56) for $\theta$.
Now, let us do the the same derivation for${{\delta}_{SUSY}}X^{\mu}$.
There is a subtle point containing in this derivation.That is, since the
product Q and X is non-singular, the result for
${{\delta_{SUSY}}X^{\mu}}$ should formally be zero.However,
let us consider the supersymmetry generator in the different picture:
\eqn\lowen{{{\Gamma}_{1}}Q={\epsilon^{A}}{\oint}{{dz}\over{2i\pi}}exp[1/2\phi]{\Sigma_{B}}{\partial}{X^{\nu}}{\gamma_{\nu}^{AB}}}
Then
%
\eqn\lowen{{{\Gamma}_{1}}{{\delta}_{SUSY}}X^{\mu}=i{\epsilon_{A}}{{\Sigma}_{B}}{\gamma^{\mu}_{AB}}exp[1/2\phi]=i\epsilon{\Gamma^{\mu}}\theta.}
Therefore, up to picture changing-transformation, the supersymmetry
transformation (56) for $X^{\mu}$ is also fulfilled.
Therefore, we find that the space-time supersymmetry (56) admits the
substitution (59).
Now we have to check the equations of motion (58).

Let us first  the quantity $\Pi_{\mu}$ in terms of NSR
variables.
Rather then doing it straightforwardly through (55) and (59) we will
compute it  in the $-1$- picture which will allow us to clearly
see the connection with with canonical relations.
The conjugate momentum
$\pi_{\theta}^B$ should satisfy the following
canonical relation:
\eqn\lowen{[\theta^{A
},{\pi_{\theta}^B}]=-{2\over{i\pi}}{\epsilon^{AB}},}
The operator that satisfies this property is
\eqn\lowen{{\pi_{\theta}^{A}}={i\over{\pi}}{\pi}exp[1/2\phi]{\Sigma^{A}}.}
Indeed, by evaluating its O.P.E with $\theta$ and taking the
most singular term we have
\eqn\lowen{:-i{\pi}exp[1/2\phi]{\Sigma^{A}}:(z):exp[-1/2\phi]{\Sigma^{B}}:(w)
{\sim}{{1\over{-i{\pi}}}}{{\epsilon^{AB}}\over{z-w}}+...}
On the other hand,by variating the action (54) we have
\eqn\lowen{{\pi_{\theta}^{A}}={i\over{\pi}}{\gamma^{\mu}}{\Pi_{\mu}}{\theta_{A}},}
therefore we find
\eqn\lowen{{\Pi_{\mu}}=2exp[-\phi]{\psi_{\mu}}}
since this operator satisfies (67).
This result is,of course, in accordance (up to picture-changing) with
the definition (55) of $\Pi^{\mu}$ since
\eqn\lowen{{\partial}X^{\mu}-i\theta{\gamma^{\mu}}\partial\theta=({\Gamma_{1}}+{{\Gamma}_{2}})exp[-\phi]{\psi^{\mu}}+{\lbrace}Q_{BRST},...{\rbrace}}
Given this, let us next calculate the product
\eqn\lowen{T=1/2{\Gamma_{2}}{\Pi^{\mu}}{\Pi_{\mu}}}
which is the picture-changed stress energy tensor for the GS superstring .
It was shown in~\refs{\berkk} that  there exists a rather complicated
field redifinition~\refs{\berkk, \berk} that transforms the N=2  stress-energy
tensor of  the critical GS supersting in a Calabi-Yau background (with 6
compactified dimensions)
into a twisted N=2 tensor which is constructed from a combination
of NSR ghosts and a shifted BRST current~\refs{\vafa}.We will show
how, by using the  redefinition (59) to construct (up to picture-changing)
an untwisted NSR
stress-energy tensorout of the N=1 GS stress-energy tensor of the theory
 in a flat background (70).The importance of such a construction is
that it is relevant to the question of quantization of the GS
superstring in a flat background.The difficulties of such the
quantization were discussed in~\refs{\berkk} and are related to the fact
that $\theta^A=exp[1/2\phi]{\Sigma^A},A=1,...16$ are not free fields.
We have:
%
\eqn\grav{\eqalign{:{\Pi_{\mu}}{\Pi^{\mu}}:=:exp[-\phi]{\psi_{\mu}}exp[-\phi]{\psi^{\mu}}:=exp[-2\phi][4\partial{\psi_{\mu}}{\psi^{\mu}}+{\partial^{2}}\phi-\partial\phi\partial\phi];}}
then
\eqn\grav{\eqalign{:{\Gamma_{1}}{\Pi_{\mu}}{\Pi^{\mu}}:
=:[-1/2exp[\phi]{\psi_{\nu}}{\partial}X^{\nu}-1/2exp[2\phi-\kappa]\partial{\phi}b+exp[\kappa]\partial{\kappa}c]{\times}\cr
{\times}exp[-2\phi](4\partial{\psi_{\mu}}{\psi^{\mu}}+{\partial^2}\phi-\partial\phi\partial\phi):=\cr
=2exp[-\phi]{\psi_{\nu}}\partial{X^{\nu}}+exp[\kappa-2\phi]\partial{\kappa}c[4\partial{\psi_{\mu}}{\psi^{\mu}}+{\partial^2}\phi-\partial\phi\partial\phi]}}
Finally,the product of this expression with $\Gamma_1$ gives:
%
\eqn\grav{\eqalign{:{\Gamma_{2}}{\Pi_{\mu}}{\Pi^\mu}:=:{\Gamma_1}{\Gamma_1}{\Pi_\mu}{\Pi^\mu}:=\cr
=\partial{X_{\mu}}\partial{X^\mu}+\partial{\psi_{\mu}}{\psi^\mu}+\partial\sigma\partial\sigma+3{\partial^2}\sigma+\cr
+\partial\kappa\partial\kappa+{\partial^2}\kappa-\partial\phi\partial\phi-2{\partial^2}\phi.}}
Here, as previously, $\sigma$ is a bosonized fermionic ghost while
the fields $\phi$and $\kappa$ come from the bosonization formulas (21)
for $\beta$,$\gamma$ bosonic ghosts.
The $\sigma$-term in (73)
\eqn\lowen{2T_{b-c}=\partial\sigma\partial\sigma+3{\partial^2}\sigma}
is exactly twice the stress-energy tensor of the fermionic ghost system
while the term
\eqn\lowen{2T_{\beta-\gamma}=-\partial\phi\partial\phi-2{\partial^2}\phi+\partial\kappa\partial\kappa+{\partial^2}\kappa}
is the stress-energy tensor for the $\beta-\gamma$ ghost system written
in the bosonized form.
Therefore we find that
\eqn\lowen{1/2:{\Gamma_{2}}{\Pi_\mu}{\Pi^\mu}:=T_{matter}+T_{b-c}+T_{\beta-\gamma}}
But,up to picture-changing transformation,$1/2{\Pi_\mu}{\Pi^\mu}$ is
a stress-energy tensor for a Green-Schwarz superstring.
Therefore by using the equivalence relation (59),we  have found
the exact relation between GS and NSR stress-energy tensors to be given by
\eqn\lowen{{\Gamma_2}T_{GS}=1/2{({\Gamma_{1}}+{\Gamma_{2}})^2}T_{NSR}}
This very important result means that the proposed
the equivalence relation  (59) between GS and NSR observables
transforms ,up to picture-changing,$T_{GS}$-the stress-energy
tensor of Green-Schwarz superstring to $T_{NSR}$-of NSR superstring.
Let us now check that in NSR formalism the equations of motion (58)
are still fulfilled.
Let us check the first equation in (58).
In the NSR formalism the fulfillment of the
equation ${\gamma^\mu}{\Pi_\mu}{\bar{\partial}}\theta=0$ means
that all the terms of the O.P.E. implied in this equation vanish.
This is true if and only if the product of $\Pi(z)$ with $\theta(w)$ does not
contain the term proportional to $\sim$ $1\over{z-w}$.
Since
${\gamma^\mu}{\Pi_\mu}(z)\theta(w)={{(z-w)}^0}exp[-1/2\phi]{\Sigma}+...$
we see that the first equation in (58) is indeed fulfilled in the NSR
formalism.
The second equation of motion in (58) is satisfied because
\eqn\grav{\eqalign{{\bar{\partial}}(\partial{X^\mu}+i\theta{\gamma^\mu}{\partial}{\theta})={\bar{\partial}}({\Gamma_1}{\Pi^\mu}-c\beta{\psi^\mu})=0}}
since the product ${\Gamma_1}(z){\Pi^\mu}(w)$ does not contain singular
terms.
Therefore the equations of motion (58) are satisfied in the NSR
formalism.
Now, we have to show that the relation (59) leaves the
equations of motion (58) invariant under the $\kappa$-symmetry
transformations (57).
We find that
the variation of the first equation in
(58) under the transformations (57) is given by
\eqn\lowen{{\delta_{\kappa}}({\gamma^\mu_{AB}}{\Pi_\mu}{\bar{\partial}}{\theta^B})=-2{\partial\theta_C}{\kappa^C}{\gamma^\mu_{AB}}{\Pi_\mu}{\bar{\partial}}{\theta^B}+i{\kappa_A}{\Pi^\mu}{\bar{\partial}}{\Pi_\mu}}
The first term is zero because of the equations of motion while the
second one vanishes because the product expansion of
${\Pi^\mu}(z)$ with ${\Pi_\mu}(w)$ contains no term proportional
to ${\sim}{1\over{z-w}}$.The $\kappa$-variation of the second equation
of motion gives:
\eqn\lowen{{\delta_\kappa}{\bar{\partial}}{\Pi^\mu_z}={\bar{\partial}}{\delta_\kappa}{\Pi^\mu_z}=
-4{{\bar{\partial}}_{\bar{z}}}{{lim}_{z{\rightarrow}w}}{\partial}\theta(z){\gamma^\mu}{\gamma_\nu}{\Pi^\nu_w}\kappa.}
Again,in order for this variation to vanish,
the product $\partial\theta(z){\Pi^\nu}(w)$ should contain no term
proportional to ${\sim}{1\over{z-w}}$.Writing this product in terms of
the NSR formalism (with $\Pi$ written in the $-1$- picture) and

leaving the most singular terms (which are proportional to ${1\over{z-w}}$):
\eqn\grav{\eqalign{\partial\theta(z){\Pi^\nu}(w)=(1/2\partial{\phi}exp[1/2\phi]\Sigma(z)exp[-\phi]{\psi^\nu}(w)+exp[1/2\phi]\partial\Sigma(z)exp[-\phi]{\psi^\nu}(w))=\cr
=1/2{1\over{z-w}}{\gamma^\nu}exp[-1/2\phi]{\Sigma}(w)-1/2{1\over{z-w}}{\gamma^\nu}exp[-1/2\phi]{\Sigma}(w)+...=0\times{1\over{z-w}}+...}}
Therefore, we find that,in the NSR formalism the equations of motion of the GS
superstring theory do not change under $\kappa$-transformation.
As a result, we  find that the formula (59) that relates GS and NSR
parameters indeed establishes the "mapping" between GS and NSR
superstring theories, up to picture-changing transformations.
The first very important consequence of this fact is that now
we can write down the explicit form of the target space supersymmetry
for the NSR superstring theory.Repeating the arguments above we have:
\eqn\grav{\eqalign{{\delta_{SUSY}}(exp[1/2\phi]\Sigma_{A})={\epsilon_{A}}\cr
{\Gamma_{1}}{\delta_{SUSY}}X_{\mu}=i\epsilon_A{\gamma_\mu^{AB}}exp[1/2\phi]{\Sigma_B}.}}
These are space-time supersymmetry transformations for  NSR superstring theory.
 We have to stress that, while the relation (59) always allows us
to express GS variables in terms of NSR variables,the reverse (NSR in
terms of GS) is possible only for GSO-projected NSR operators~\refs{\vaffa}
Let us now return to $\kappa$-transformations and their relation to
NSR formalism.
The physical meaning of these transformations is that they remove 8
spinor components of $\theta$,thus insuring the supersymmetry.
It is therefore reasonable to expect that in NSR formalism they would
correspond to BRST symmetry, which has the similar function.
In NSR formalism,$\kappa$-transformations will look  in conformal
gauge as follows:
%
\eqn\grav{\eqalign{{\delta_\kappa}(exp[1/2\phi])\Sigma)=2i{\gamma^\mu}exp[-\phi]{\psi_\mu}\kappa\cr
{\Gamma_1}{\delta_\kappa}{X^\mu}=-2{\gamma^\mu}exp[-1/2\phi]\Sigma\kappa}}
In order to investigate how $\kappa$-transformation acts on
an arbitrary operator in NSR formalism it is useful to  write down
the generator of $\kappa$-transformations.From (83) we deduce that
in  NSR formalism this generator is (up to the picture-changing)
\eqn\lowen{{G_{\theta}^A}=exp[-3/2\phi]{\Sigma^A}=V_{-3/2}(0)}
Using this formula, it is now easy to check that
$\kappa$-transformation of the picture-changing operator $\Gamma_1$
give:
\eqn\lowen{{\delta_\kappa}{\Gamma_1}=exp[\kappa-3/2\phi(z)]c(z){\partial\kappa(z)}{\Sigma^A}{\kappa_{A}}}
If case when $\kappa$-transformation is applied to vertex operators,the
term associated with the variation of ${\Gamma_1}$ gives the
contribution proportional to the ghost field $c(z)$ and therefore
can be omitted since it does not contribute to correlation functions.

Let us examine closer how $\kappa$-transformation works for vertex
operators.Let us first of all consider
$V_{1/2}(k=0)=\oint{{dz}\over{2i\pi}}V_{1/2}(k=0)$.We have:
%
\eqn\grav{\eqalign{{\delta_\kappa}V_{1/2}(k=0)={\delta_{\kappa}}\oint{{dz}\over{2i\pi}}{\Gamma_1}\theta{\gamma^\mu}{\Pi_{\mu}}={\Gamma_{1}}{\oint}{{dz}\over{2i\pi}}{\gamma^\mu}{\Pi_\mu}{\delta_\kappa}\theta=\cr
=2i{\Gamma_{1}}{\oint}{{dz}\over{2i\pi}}{\Pi_\mu}{\Pi^\mu};\cr
{\Gamma_{1}}{\delta_\kappa}V_{1/2}(k=0)=2i{\lbrace}Q_{BRST},{\oint}{{dz\over{2i\pi}}}b(z){\rbrace}}}
The same result can, of course, be obtained by directly applying the
generator (84) to $V_{1/2}$ and making picture-changing manipulations.

We see that $\kappa$-transformation leaves zero momentum fermionic
vertices invariant up to BRST trivial terms.
It is interesting to notice that the generator (84) resembles
 the space-time supersymmetry current which is given by
$j_{SUSY}=exp[-1/2\phi]\Sigma=V_{-1/2}(z,k=0)$.
It is because ${\Gamma_1}V_{-3/2}(k=0)=0{\neq}V_{-1/2}(k=0)$ that the
generator of the $\kappa$-symmetry is not related to the supersymmetry
current through picture-changing transformation and therefore the
$\kappa$-symmetry leads to some new identities between NSR scattering
amplitudes which will be discussed below.

Using (83) let us now compute the $\theta$-transformation of  bosonic
 vertices with zero momenta.To be certain,let us take $V^\mu_{0}(z,k=0)$.
Omitting terms that depend only on $c(z)$ and do not contribute
to correlation functions  we have:
\eqn\grav{\eqalign{0={\delta_\kappa}V^{\mu}_{0}(z,k=0)={\delta_\kappa}{\partial}{X^\mu}(z)={\delta_\kappa}({\Gamma_1}exp[-\phi]{\psi^\mu}(z)+c(z)\beta{\psi^\mu}){\sim}{\Gamma_1}{\delta_\kappa}(exp[-\phi]{\psi^\mu})=\cr
={\Gamma_1}{\delta_\kappa}{\Pi^\mu}={\Gamma_1}2i\partial\theta{\gamma^\mu}{\times}2i{\Gamma^\nu}{\Pi_{\nu}}}}
as a consequence of the equations of motion.
$\kappa$-transformations for vertices in other pictures is obtained by
picture-changing transformation.
Therefore we now see that bosonic and fermionic vertex operators are
invariant under $\kappa$-transformation.
The situation is different  for vertex operators with non-zero momenta
which are no longer $\kappa$-invariant.
Let us compute , for example, ${\delta_\kappa}V_{0}(k,z)$.After
rather cumbersome computations we get:
\eqn\grav{\eqalign{{\delta_\kappa}V_{-1}(k,z)={\Gamma_1}{\delta_\kappa}{e^\mu}(k){\Pi_\mu}exp[ikX]={\Gamma_1}{e^\mu}(k){\Pi_\mu}{\delta_\kappa}exp[ikX]=\cr
=:{\Gamma_1}{e^\mu}(k){\Pi_\mu}{\kappa^A}exp[-3/2\phi]{\Sigma_A}exp[ikX]:=\cr
=i/2{{e^\mu}}(k){\kappa^A}k_{\alpha}{{[{\gamma^\mu},{\gamma_\alpha}]}_{AB}}{\partial}[exp[-3/2\phi]{\Sigma^B}exp[ikX]]={\tilde{\bar{u^B}}}(k){\partial}V_{{-3/2}{B}}(z,k),}}
where
${\tilde{\bar{u^B}}}(k)=i/2{e_\mu}(k){k_{\alpha}}{\kappa^A}{{[{\gamma^\mu},{\gamma^\alpha}]}_{AB}}$
is the "effective spinor" which determines the spirality.
Similarly,evaluation of the $\kappa$-transformation of the fermionic
vertex ${\delta_\kappa}V_{1/2}(k,z)$ gives
\eqn\lowen{{\delta_\kappa}{\gamma^\mu_{AB}}exp[1/2\phi]{\Sigma^B}({\partial}{X_\mu}+i/4(k\psi){\psi_\mu})exp[ikX]=i/2{\partial}{\tilde{e_\mu}}(k){V^\mu_{-1}}(z,k),}
where
\eqn\lowen{{\tilde{e_\mu}}(k)=i/2{\kappa^A}{{({\gamma^\alpha}{\gamma^\nu}{\gamma_\mu}{\gamma_\nu})}_{AB}}{k_\alpha}{\bar{u_B}}.}
Thus, at non-zero momenta $\kappa$-symmetry transforms bosonic vertex
operators into derivatives of fermionic ones, and vice versa for
fermionic vertices.
This shows that, as we should have expected,string amplitudes are
independent under $\kappa$-transformations;indeed,to obtain a scattering
amplitude we would have to integrate each vertex operator $V(z_{i})$ over
$z_{i},i=1,...N$; and the integrals of full derivatives would give zero.
As of unintegrated correlation functions, $\kappa$-symmetry may give
new relations between bosonic and fermionic amplitudes additional to
those following from supersymmetry.Consider, for example,the correlation
function in which all the momenta except for the first two are zero;
let us find its theta transformation.For the sake of certainty, let us
the first particle be a boson and the second one a fermion.
We have:
\eqn\grav{\eqalign{{\delta\kappa}<V_{B}({z_1},k)V_{F}(z_{2},-k)V(z_{3},k=0)...V(z_{N},k=0)>=\cr
=<{\delta_\kappa}V_{B}(z_{1},k){V_F}(z_{2},-k)V(z_{3},k=0)...V(z_{N},k=0)>+\cr
+<V_{B}(z_{1},k){\delta_\kappa}{V_F}(z_{2},-k)V(z_{3},k=0)...V({z_N},k=0)>=\cr
={\partial_{z_{1}}}<V_{F}V_{F}...>+{\partial_{z_{2}}}<V_{B}V_{B}...>=0.}}
The principal distinction of $\kappa$-symmetry from the supersymmetry
is that the first one does distinguish between zero and non-zero momenta
while the latter does not.

\centerline{\bf 8.Conclusion}
We have shown that the appearance of picture-changing operators in the
String Theory is the consequence of imposing various gauge constraints
on the symmetry group (2).The mechanism of the picture-changing is derived
from the first principles,i.e. by computing the Faddeev-Popov
determinant with the gauge constraints (3).While the similar computations
have already been performed in~\refs{\ver} for $\Gamma_1$, the
computation for higher picture-changing operators has been done for the
first time.The exact expression for $\Gamma_n$ has been obtained.
We proved that, up to BRST cohomology  various $\Gamma_n$'s form
the polynomial ring (44).Using this we show that the
the correlators are independent on  pictures  in which the vertex operators.
are taken.
It is not clear yet if it is possible
to extend the algebra (44) to half-integer
values of $n$.
 Using the properties of picture-changing,we have
found the correct formula (52) for the fermionic perturbations of
a background.This should allow us to compute the $\beta$-function of
gravitino; hopefully that can be done in  future papers.
 Furthermore, we have established a relation between the $N=1$
GS supersring theory in a flat background and
and the N=1 NSR superstring theory
 and have shown that the supersymmetry
transformations,equations of motion
 and $\kappa$-invariance of the GS superstring theory to be
properly fulfilled in the NSR formalism.We have shown that the N=1 GS
stress-energy tensor (of the theory in a flat background)
is transformed into the stress-energy tensor of the
N=1 NSR superstring theory.
The identities of GS Superstring Theory were shown to be fulfilled in the
NSR formalism in terms of operator product expansions.
This connection allows to solve solve the problem of the covariant
quantization of GS Superstring theory by reducing it to NSR formalism
for which the covariant quantization is well-known.
As a consequence of this relation, we were able to derive the target
space supersymmetry transformations (82).
These transformation laws include picture-changing operators
in a natural way.
Many interesting questions still remain in connection with
$\kappa$-symmetry;establishing its relation to BRST  would be
especially important.For bosonic particles, BRST symmetry is related to
the fact that we can add to the polarization vector the quantity
proportional to the momentum of the particle ${\sim}k^\mu$.The fact that
$\kappa$-transformations applied to vertex operators give
new effective polarization vectors (spinors for fermions) which are
also proportional to the momentum of the particle and the fact that
both $\kappa$-symmetry and BRST have the similar function of removing
redundant states - both these facts show that such kind of relation
should exist.
 Finally, it was shown that the
$\kappa$-symmetry allows us do derive the
relations between bosonic and fermionic correlation functions,additional
to the relations following from the space-time supersymmetry.
\centerline{\bf Acknowledgements}
This work was supported from the Grant of the String Theory
group of Rutgers University.Many thanks to
D.Friedan,S.Shenker and A.B.Zamolodchikov for their helpful commentaries.
 I'm  especially obliged and grateful to A.M.Polyakov for his
suggestions.
%

\vskip .5in
\footatend\vfill\immediate\closeout\rfile\writestoppt
\centerline{\bf{References}}\bigskip{\frenchspacing%
\parindent=20pt\escapechar=` 
\input \jobname.refs\vfill\eject}\nonfrenchspacing

\end